\documentclass[twocolumn]{revtex4}               
\usepackage{amsmath,amssymb}
\usepackage{latexsym}
\usepackage{epsfig}

\begin{document}
\title{Tsallis Holographic Dark Energy in the Brans-Dicke Cosmology}
\author{S. Ghaffari$^{1}$\footnote{sh.ghaffari@riaam.ac.ir}, H. Moradpour$^1$\footnote{hn.moradpour@gmail.com}, I. P.
Lobo$^{2}$\footnote{iarley\_lobo@fisica.ufpb.br}, J. P.
Morais Gra\c ca$^{2}$\footnote{jpmorais@gmail.com}, Valdir B. Bezerra$^{2}$\footnote{valdir@fisica.ufpb.br}}
\address{$^1$ Research Institute for Astronomy and Astrophysics of Maragha (RIAAM), P. O. Box 55134-441, Maragha, Iran\\
$^{2}$ Departamento de F\'{i}sica, Universidade Federal da Para\'{i}ba, Caixa Postal 5008, CEP 58051-970, Jo\~{a}o Pessoa, PB, Brazil}

\begin{abstract}
Using the Tsallis generalized entropy, holographic hypothesis and
also considering the Hubble horizon as the IR cutoff, we build a
holographic model for dark energy and study its cosmological
consequences in the Brans-Dicke framework. At first, we focus on a
non-interacting universe, and thereinafter, we study the results of
considering a sign-changeable interaction between the dark sectors
of the cosmos. Our investigations show that, compared with the flat
case, the power and freedom of the model in describing the cosmic
evolution is significantly increased in the presence of the
curvature. The stability analysis also indicates that, independent
of the universe curvature, both the interacting and non-interacting
cases are classically unstable. In fact, both the classical
stability criterion and an acceptable behavior for the cosmos
quantities, including the deceleration and density parameters as
well as the equation of state, are not simultaneously obtainable.
\end{abstract}
 \maketitle

\section{Introduction}

Cohen et al's proposal \cite{HDE} gives us an estimation for the
upper bound of the energy density of quantum fields in the vacuum
states. Shortly afterwards, it has been proposed that this bound may
provide an explanation for dark energy (DE), a hypothesis called
Holographic dark energy (HDE), is a promising approach to solve the
dark energy problem, and its related topics
\cite{HDE5,HDE17,HDE01,HDE1,HDE2,HDE3,stab,RevH,wang}. Indeed, the
HDE hypothesis helps us in finding the cosmological features of the
vacuum energy. The mutual relation between the UV and IR cutoffs
forms the backbone of HDE \cite{RevH,wang}. Finally, it is
worthwhile to mention here that any changes in the horizon entropy,
including changes in $i$) the entropy-area relation, $ii$) the IR
cutoff or even a combination of these ways, lead to new models for
HDE \cite{RevH,wang,Sayahian,Renyi,Tavayef}.

The Brans-Dicke (BD) theory of gravity is an alternative to general
relativity in which the gravitational constant $G$ is not constant,
and it is replaced with the inverse of a scalar field ($\phi$)
\cite{4not}. Although the BD theory can provide a description for
the current universe \cite{bdde}, its theoretical predictions for
the $w$ parameter has major difference with the observations
\cite{jcap,jcap1,pav}. In fact, the theoretical estimations for the
value of $w$ is much less than those are obtained from observations,
a result encouraging physicists to use various dark energy sources
in order to describe the current universe in the BD framework
\cite{jcap,jcap1,pav}.

Motivated by the above arguments, the idea of HDE has also been
employed to study the dark energy problem in the BD framework
\cite{Gong,Setare,Banerjee2,Banerjee,Xu,Jamil,Khodam}. It has also
been argued that since HDE is a dynamic model, one should use the
dynamic frameworks, such as the BD theory, to study its cosmological
features \cite{Setare,Jamil}. It has been shown that the original
HDE with the Hubble radius as IR cutoff cannot produce the cosmic
acceleration in the BD theory \cite{Xu}, while for the event horizon
as the IR cut-off, an accelerated universe is obtainable.
Furthermore, it has been demonstrated that when an interaction
between HDE and DM is taken into account, the phantom line is
crossed in the BD cosmology \cite{Jamil}. The stability of
interacting HDE with the GO cutoff in the BD theory has also been
discussed in \cite{Khodam}. Observations also admit a
sign-changeable interaction between the cosmos sectors
\cite{Wei,Chimento1,Chimento2}. Such interaction usually admits the
cosmological models to experience a phantom phase \cite{Abdolahi}.

Recently, using Tsallis generalized entropy \cite{Tsallis}, and by
considering the Hubble horizon as the IR cutoff, in agreement with
the thermodynamics considerations \cite{Sayahian,Renyi}, a new HDE
model, called Tsallis Holographic Dark Energy (THDE), has been
introduced and studied in the standard cosmology framework
\cite{Tavayef}. At first glance, it is a proper model for the
current universe in the standard cosmology framework
\cite{Tavayef,maj,sar}, but, the same as the primary HDE based on
the Bekenstein entropy \cite{stab}, THDE is not stable
\cite{Tavayef,maj,sar}. More studies on the various cosmological
features of the Tsallis generalized statistical mechanics can be
found in Refs. \cite{n1,n2,n3,n4}. It is also useful to note here
that a non sign-changeable interaction between the cosmos sectors
can not bring stability for this model \cite{maj}.

Here, we are interested in studying the consequences of employing
the THDE model in modeling dark energy in the BD cosmology. In our
setup, the Hubble horizon as the IR cutoff is taken into account,
and both the interacting and non-interacting cases are also
investigated. In order to achieve this goal, we studied the
non-interacting case in the next section. The situation in which
there is a sing-changeable interaction between the cosmos sectors
has also been addressed in Sec.~($\textmd{III})$. The fourth section
includes our results about the classical stability of the obtained
models against perturbations. The last section is devoted to
concluding remarks.

\section{Non-interacting Tsallis holographic dark energy in the Brans-Dicke cosmology}

We consider a homogeneous and isotropic FRW universe described by the line element
\begin{equation}
{\rm d}s^2=-{\rm d}t^2+a^2(t)\left(\frac{{\rm d}r^2}{1-kr^2}+r^2{\rm
d}\Omega^2\right),\label{metric}
\end{equation}
where $k=0,1,-1$ represent a flat, closed and open universes,
respectively. For \textbf{the} universe filled by a pressureless
dark matter (DM) with energy density $\rho_m$, and a DE candidate
with energy density $\rho_D$, the Brans-Dicke field equations are
found as \cite{Banerjee}

\begin{equation}\label{Friedeq01}
\frac{3}{4\omega}\phi^2\Big(H^2+\frac{k}{a^2}\Big)-\frac{\dot{\phi}^2}{2}+\frac{3}{2\omega}H\dot{\phi}\phi=\rho_M+\rho_D,
\end{equation}
\begin{equation}\label{Friedeq02}
\frac{-\phi^2}{4\omega}\Big(\frac{2\ddot{a}}{a}+H^2+\frac{k}{a^2}\Big)-\frac{1}{\omega}H\dot{\phi}\phi
-\frac{1}{2\omega}\ddot{\phi}\phi-\frac{\dot{\phi}^2}{2}\Big(1+\frac{1}{\omega}\Big)=p_D,
\end{equation}
\begin{equation}\label{motiom eq}
\ddot{\phi}+3H\dot{\phi}-\frac{3}{2\omega}\Big(\frac{\ddot{a}}{a}+H^2+\frac{k}{a^2}\Big)\phi=0.
\end{equation}
where $H=\dot{a}/a $ is the Hubble parameter and $p_D$ represents the pressure of DE.
Following \cite{Banerjee2}, we assume that the BD field $\phi$
can be described by a power function of the scale factor, namely
$\phi \propto a^n $. One can now get

\begin{equation}\label{phidot}
\dot{\phi}=nH\phi,
\end{equation}
and hence
\begin{equation}\label{phiddot}
\ddot{\phi}=n^2H^2\phi+n\dot{H}\phi,
\end{equation}
where dot denotes derivative with respect to time.

Since the Tsallis generalized entropy-area relation is independent
of the gravitational theory used to study the system \cite{Tsallis},
the energy density of Tsallis HDE (THDE) with the Hubble radius as
the IR cutoff $(L=H^{-1})$, takes the following form
\begin{equation}\label{rho1}
\rho_D=B\phi^{2\delta}H^{4-2\delta}.
\end{equation}
Here, $\phi^2=\omega/(2\pi G_{eff})$, $G_{eff}$ is the effective
gravitational constant, and we used the holographic hypothesis
\cite{HDE,HDE5,HDE17,Tavayef}. In the limiting case, where $ G_{eff}
$ is reduced to $ G $, the energy density of THDE in the standard
cosmology is restored \cite{Tavayef}. For the $ \delta=1 $ case, the
above equation also yields the standard HDE density in the BD
gravity \cite{Xu}. The dimensionless density parameters are defined
as
\begin{eqnarray}\label{Omega}
&&\Omega_m=\frac{\rho_m}{\rho_{cr}}=\frac{4\omega\rho_m}{3\phi^2H^2},\nonumber\\
&&\Omega_D=\frac{\rho_D}{\rho_{cr}}= \frac{4B\omega}{3}\phi^{2\delta-2}H^{2-2\delta},\nonumber\\
&&\Omega_\phi=\frac{\rho_\phi}{\rho_{cr}}= 2n\Big(\frac{n\omega}{3}-1\Big),\nonumber\\
&&\Omega_k=\frac{k}{a^2H^2}.
\end{eqnarray}
\begin{figure}[htp]
\begin{center}
    \includegraphics[width=8cm]{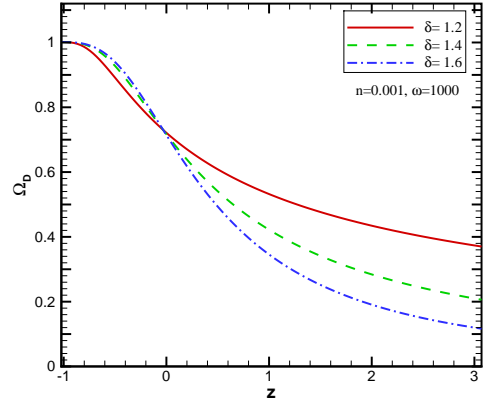}
    \includegraphics[width=8cm]{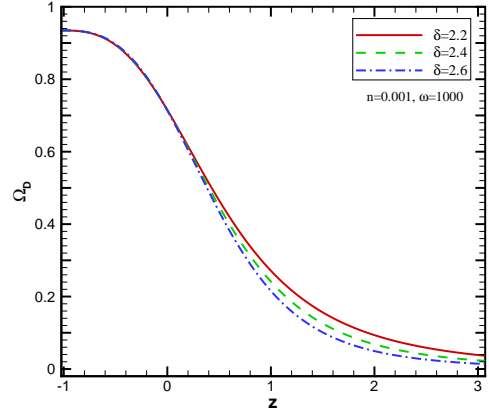}
    \caption{$\Omega_D$ versus $z$. Here, we used $ \Omega_k=0 $ (upper panel), $ \Omega_{k0}=0.1 $ (lower panel),
    $\Omega_{D0} = 0.73 $, $ n=0.001$ and $ \omega=1000 $ as the initial conditions.}\label{Omega1}
\end{center}
\end{figure}
\begin{figure}[htp]
    \begin{center}
        \includegraphics[width=8cm]{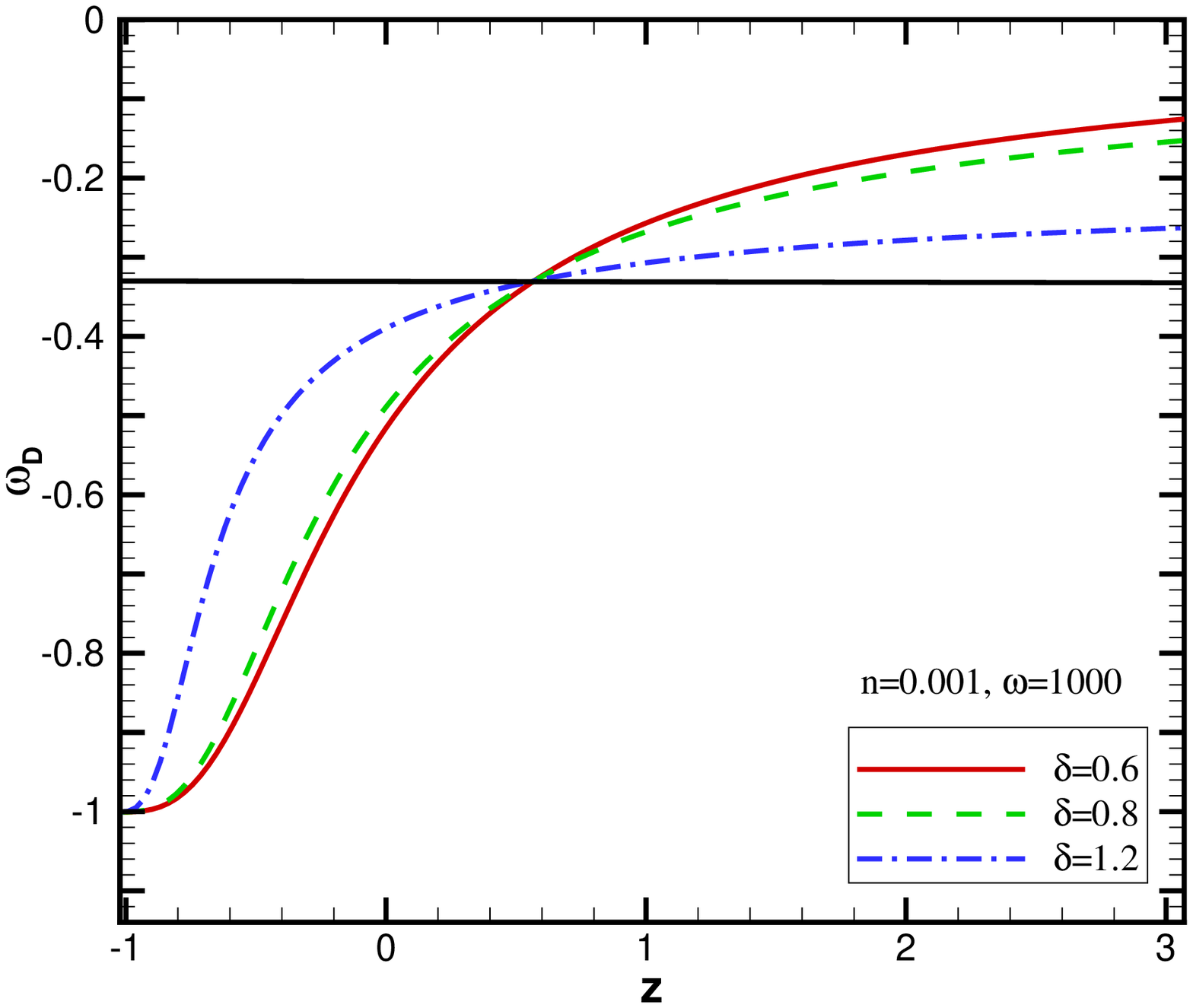}
        \includegraphics[width=8cm]{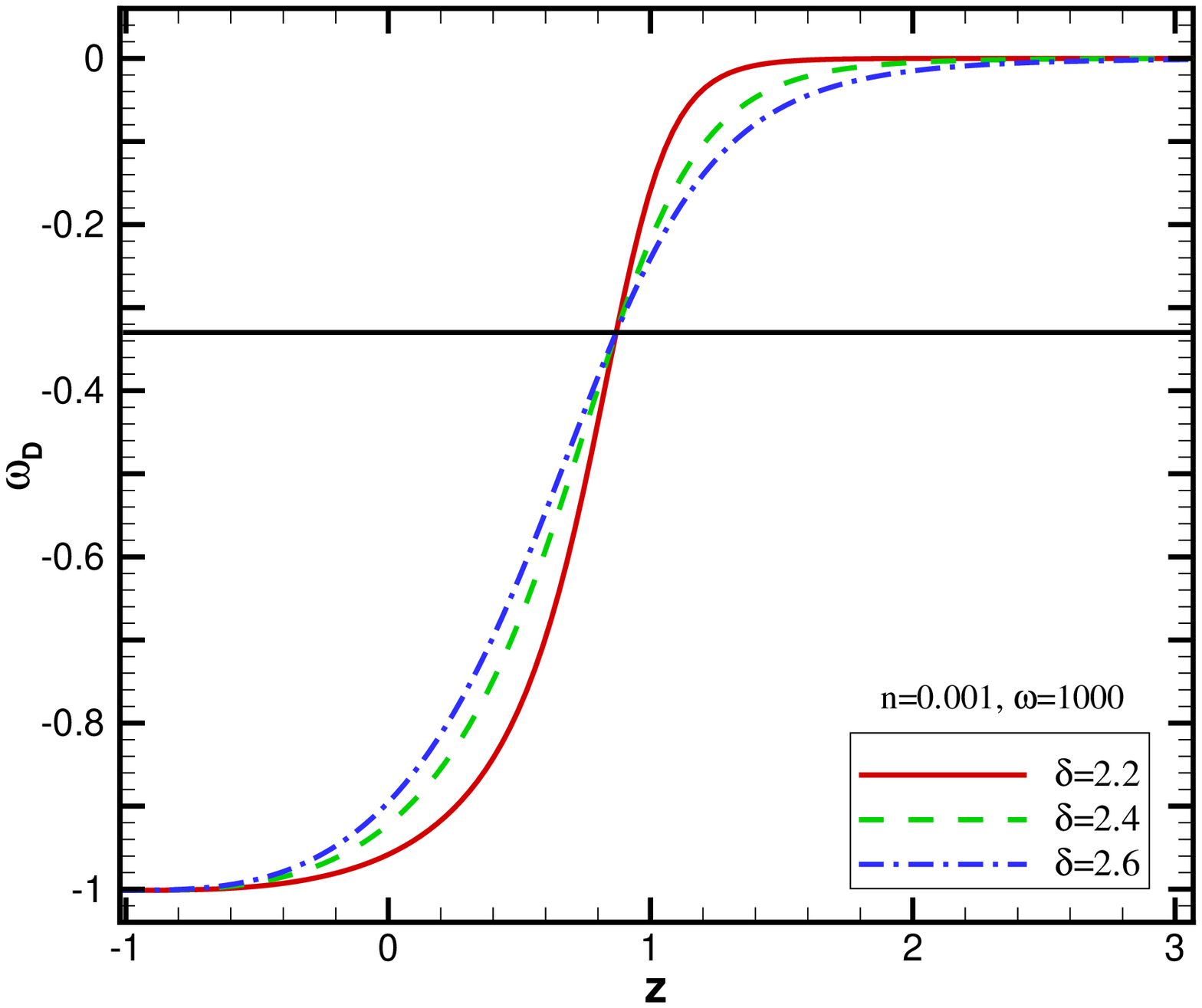}
        \caption{The evolution of the EoS parameter $\omega_D$  versus redshift parameter $z$ for the non-interacting THDE,
            where the different parameter values $\Omega_k=0 $ (upper panel), $ \Omega_{k0}=0.1 $ (lower panel),
                $ n=0.001 $ and $\omega=1000$ \cite{jcap} are adopted.}\label{figw1}
    \end{center}
\end{figure}
Here, we also assume that there is no energy exchange between the cosmos sectors, and hence, the energy conservation equations are as follows
\begin{equation}
\dot{\rho}_D+3H(1+\omega_D)\rho_D=0,\label{ConserveDE}
\end{equation}
and
\begin{equation}
\dot{\rho}_m+3H\rho_m=0,\label{ConserveCDM}
\end{equation}
where $\omega_D=\frac{p_D}{\rho_D}$ denotes the equation of state
(EoS) parameter of dark energy. Taking the time derivative of
Eq.~(\ref{rho1}), we have
\begin{equation}\label{rhodot}
\dot{\rho}_D=2H\rho_D\Big(n\delta+(2-\delta)\frac{\dot{H}}{H^2}\Big),
\end{equation}
combined with relation $\dot{\Omega}_D=H\Omega_D^\prime$ to obtain
\begin{equation}\label{dotOmega1}
\Omega_D^\prime=2(1-\delta)\Omega_D\Big(\frac{\dot{H}}{H^2}+n\Big),
\end{equation}
where prime denotes derivative with respect to $x=\ln a$. Now,
combining the time derivative of Eq. (\ref{Friedeq01}) with Eqs.
(\ref{phidot}), (\ref{phiddot}), (\ref{ConserveCDM}) and
(\ref{rhodot}), one can easily get
\begin{eqnarray}\label{Hdot1}
&&\frac{\dot{H}}{H^2}=\Big[3(\Omega_D-1)-\Omega_k+\nonumber\\&&\nonumber
2n(\delta\Omega_D+\frac{2\omega n^2}{3}+n\omega-2n-\Omega_k-4)\Big]\times\\&&
\Big(2(\delta-2)\Omega_D-\frac{4n^2\omega}{3}+4n+2\Big)^{-1}
\end{eqnarray}
\begin{figure}[htp]
\begin{center}
    \includegraphics[width=8cm]{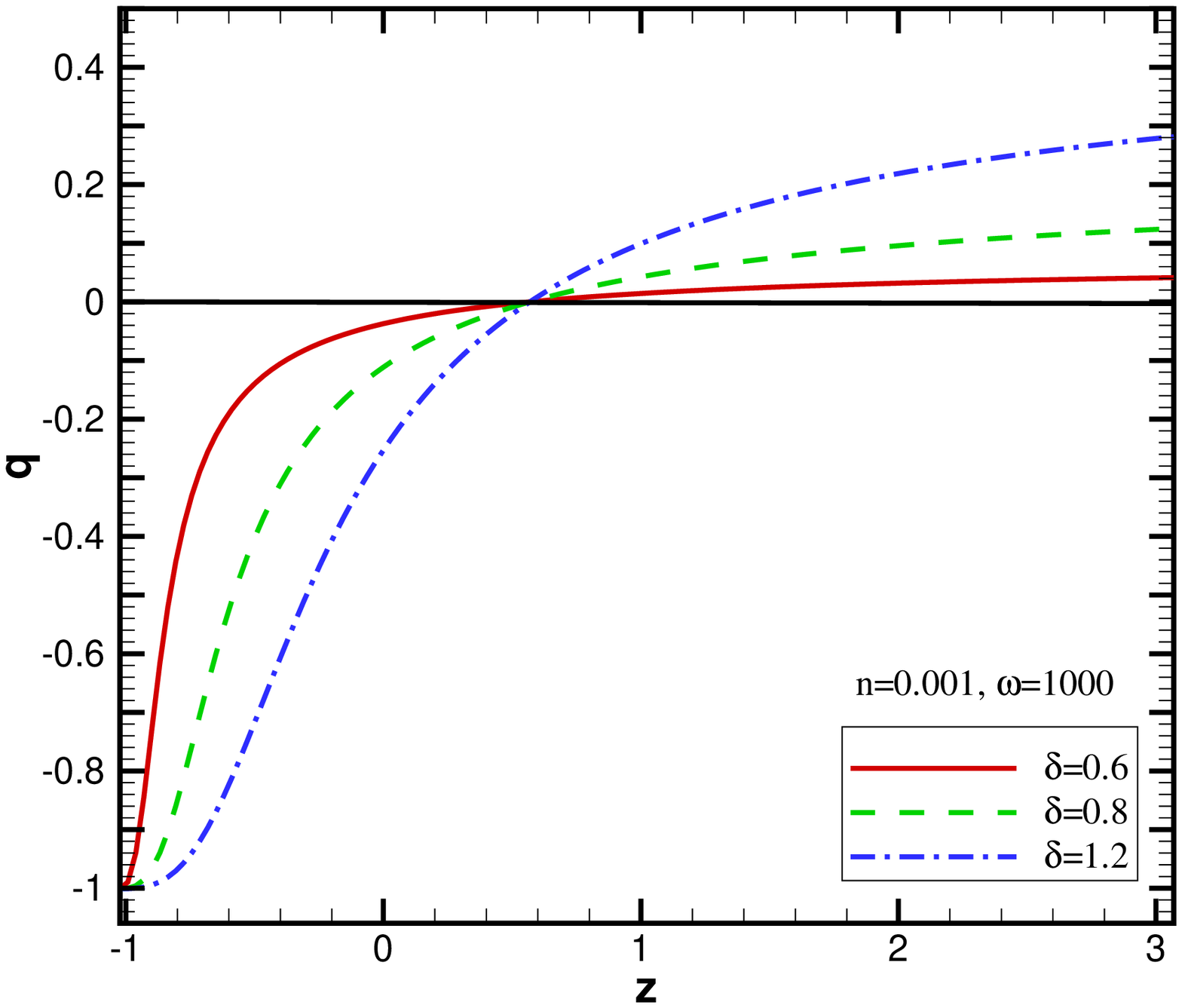}
    \includegraphics[width=8cm]{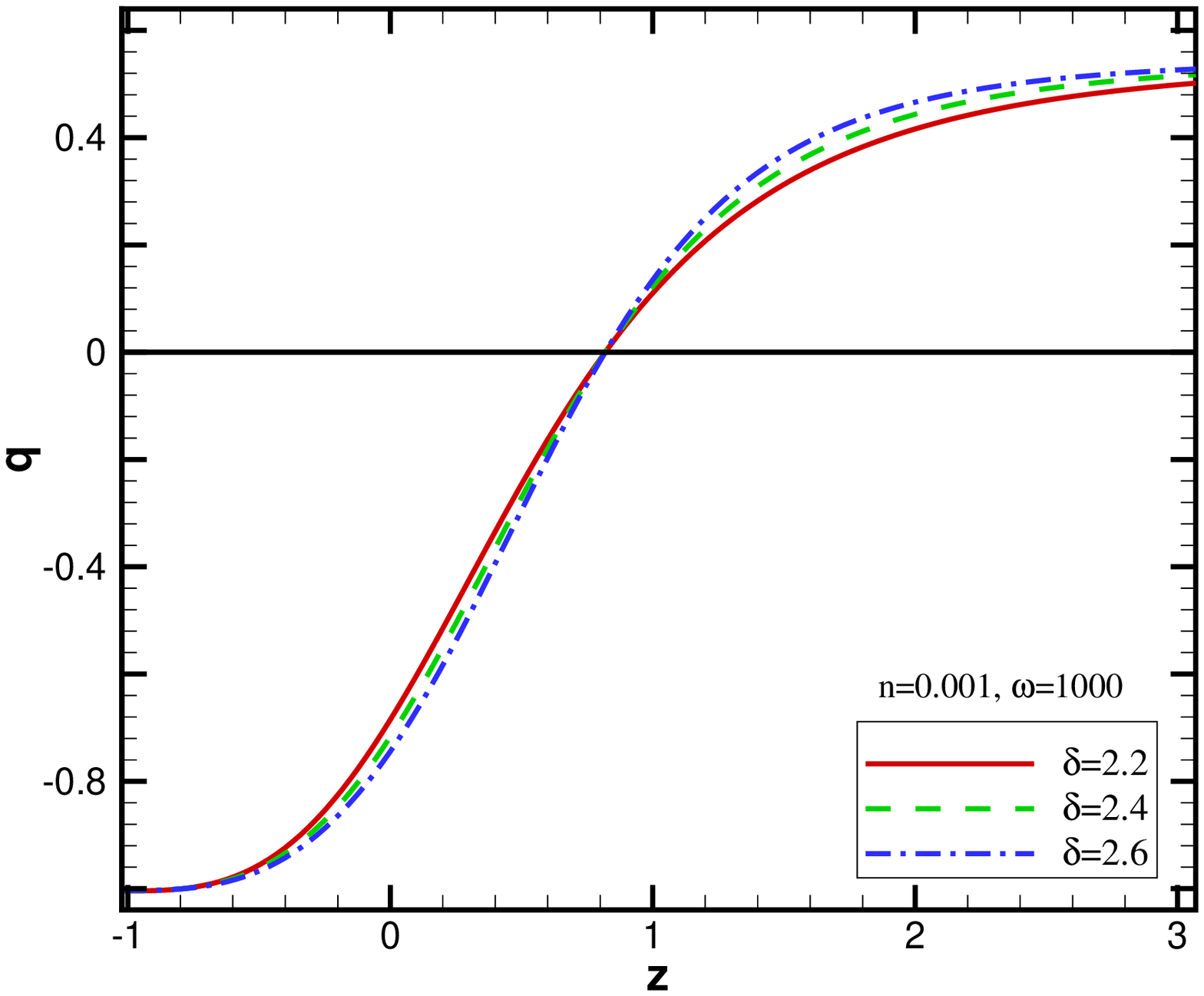}
    \caption{The evolution of the deceleration parameter $q$ versus redshift parameter $z$ for the non-interacting THDE,
    where the different parameter values $\Omega_k=0 $ (upper panel), $ \Omega_{k0}=0.1 $ (lower panel), $ n=0.001 $ and $\omega=1000$ are adopted.}\label{figq1}
\end{center}
\end{figure}
Inserting Eq. (\ref{Hdot1}) into (\ref{dotOmega1}),
we also obtain the evolution of dimensionless THDE density as
\begin{eqnarray}\label{Omegaprime1}
&&\Omega_D^\prime=2n(\delta-1)\Omega_D+\Omega_D(1-\delta)\times\\&&
\frac{\Big(3(\Omega_D-1)-\Omega_k+2n(\delta\Omega_D+n\omega+\frac{2\omega n^2}{3}-2n-\Omega_k-4)\Big)}
{(\delta-2)\Omega_D-\frac{2n^2\omega}{3}+2n+1}.\nonumber
\end{eqnarray}
In the limiting case $ n=0 $ $ (\omega\rightarrow 0) $,
Eq.~(\ref{Omegaprime1}) restores the result of the Einstein gravity
\cite{Tavayef}. The evolution of the dimensionless THDE density
parameter $\Omega_D$ against redshift $z$ is shown in Fig.
(\ref{Omega1}) for the $ \Omega_k=0 $ (the upper panel) and $
\Omega_{k0}=0.1 $ \cite{far}, where $\Omega_{k0}$ is the current
value of $\omega_k$, (the lower panel) cases whenever the initial
condition $ \Omega_D(z=0)=0.73 $ has been considered. Additionally,
$ n=0.001 $ and $\omega=1000$ \cite{jcap,jcap1} have also been used
to plot Fig. (\ref{Omega1}), showing that in the early time $
(z\rightarrow \infty) $ we have $ \Omega_D \rightarrow 0 $, while at
the late time (where $ (z\rightarrow -1) $) we have $ \Omega_D
\rightarrow 1 $. Combining Eqs. (\ref{ConserveDE}), (\ref{rhodot})
and (\ref{Hdot1}) with each other, the EoS parameter is obtained as
\begin{eqnarray}\label{w1}
&&\omega_D=-1-\frac{2\delta n}{3}+(\delta-2)\times\\&&\frac{3(\Omega_D-1)-\Omega_k+2n(\delta\Omega_D+n\omega+\frac{2\omega n^2}{3}-2n-\Omega_k-4)}
{3(\delta-2)\Omega_D-2n^2\omega+6n+3}.\nonumber
\end{eqnarray}
It is easy to see that the EoS parameter for THDE in the standard
cosmology is retrieved at the appropriate limit $ n=0 $ $
(\omega\rightarrow 0) $ \cite{Tavayef}. The behavior of $ \omega_D $
against $ z $ has been plotted in Fig. \ref{figw1}, for both the
$\Omega_k=0 $ (upper panel) and $ \Omega_{k0}=0.1 $ \cite{far}
(lower panel) cases, whenever $ n=0.001 $ and $\omega=1000$
\cite{jcap}. From Fig. \ref{figw1}, one can clearly see that the
THDE model with the Hubble cutoff in the BD gravity can lead to the
accelerated expansion, even in the absence of an interaction between
the two dark sectors of cosmos, and in addition, we have
$\omega_D(z\rightarrow-1)\rightarrow-1$ which implies that this
model simulates the cosmological constant at future.

Using Eq.~(\ref{Hdot1}), we can also write
\begin{eqnarray}\label{q1}
&&q=-1-\frac{\dot{H}}{H^2}=-1-\\&&\frac{3(\Omega_D-1)-\Omega_k+2n(\delta\Omega_D+n\omega+\frac{2\omega n^2}{3}-2n-\Omega_k-4)}
{2(\delta-2)\Omega_D-\frac{4n^2\omega}{3}+4n+2}.\nonumber
\end{eqnarray}
Once again, the respective relation in \cite{Tavayef} can be
obtained in the limiting case $ n=0 $. In the limiting case $
\delta=1 $, the obtained results in Eqs. (\ref{w1}) and (\ref{q1})
are reduced to their respective expressions for the original HDE in
the BD gravity \cite{Xu}. The evolution of $ q $ versus redshift
parameter $ z $ for different values of the parameter $ \delta $ has
also been plotted in Fig. \ref{figq1} for the $\Omega_k=0 $ (upper
panel) and $ \Omega_{k0}=0.1 $ \cite{far} (lower panel) cases,
whenever $ n=0.001 $ and $\omega=1000$ \cite{jcap}.

Our results show that the transition redshift (from the deceleration
phase to an accelerated phase) lies in the interval $ 0.5<z<0.9 $,
which is fully consistent with the recent observations
\cite{Daly,Komatsu,Salvatelli}. Figs \ref{Omega1}-\ref{figq1}
indicate that, for the assumed values of $n$ and $\omega$, $i$) only
the $\delta=1.2$ case can produce acceptable behavior for the system
quantities, including $q$, $\omega_D$ and $\Omega_D$, simultaneously
in the flat FRW universe, and $ii$) there are various values for
$\delta$ which lead to the proper behavior for the system quantities
simultaneously in the non-flat universe.

\begin{figure}[htp]
\begin{center}
    \includegraphics[width=8cm]{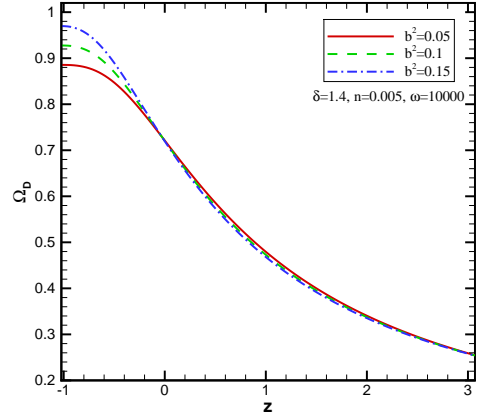}
    \includegraphics[width=8cm]{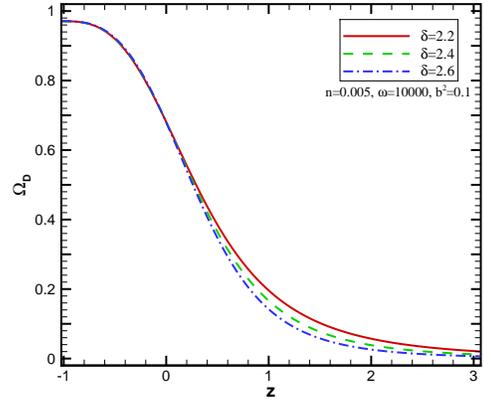}
    \caption{The evolution of $\Omega_D$ versus $z$ for the interacting THDE in the BD gravity. Here, we considered $\Omega_k=0 $ (upper panel), $ \Omega_{k0}=0.1 $ (lower panel), $\Omega_{D0} = 0.73$, $ n=0.005$ and $ \omega=10000$ \cite{jcap1} for the initial conditions.}\label{Omega2}
\end{center}
\end{figure}
\section{Sign-changeable Interacting THDE model}
\begin{figure}[htp]
\begin{center}
    \includegraphics[width=8cm]{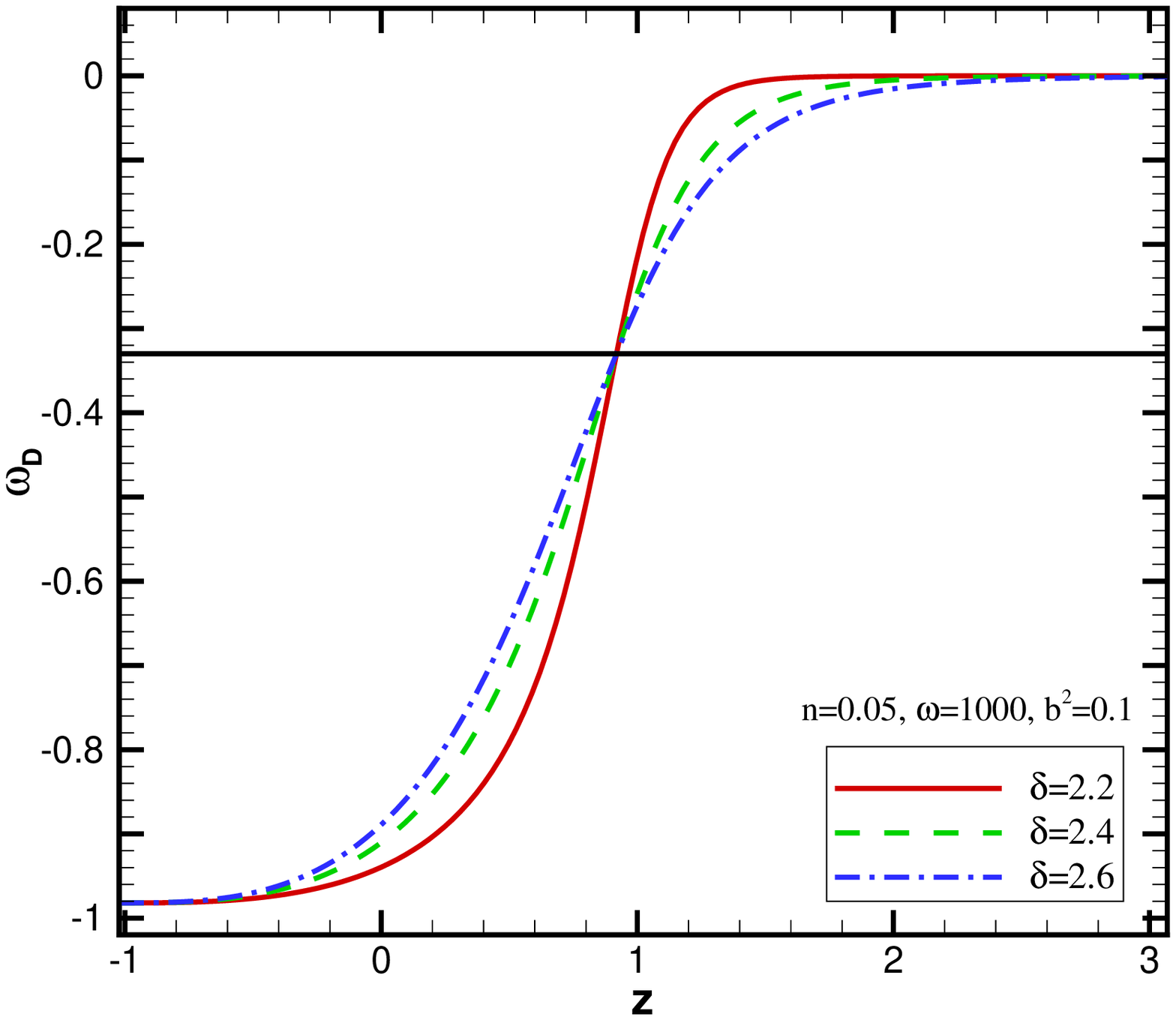}
    \includegraphics[width=8cm]{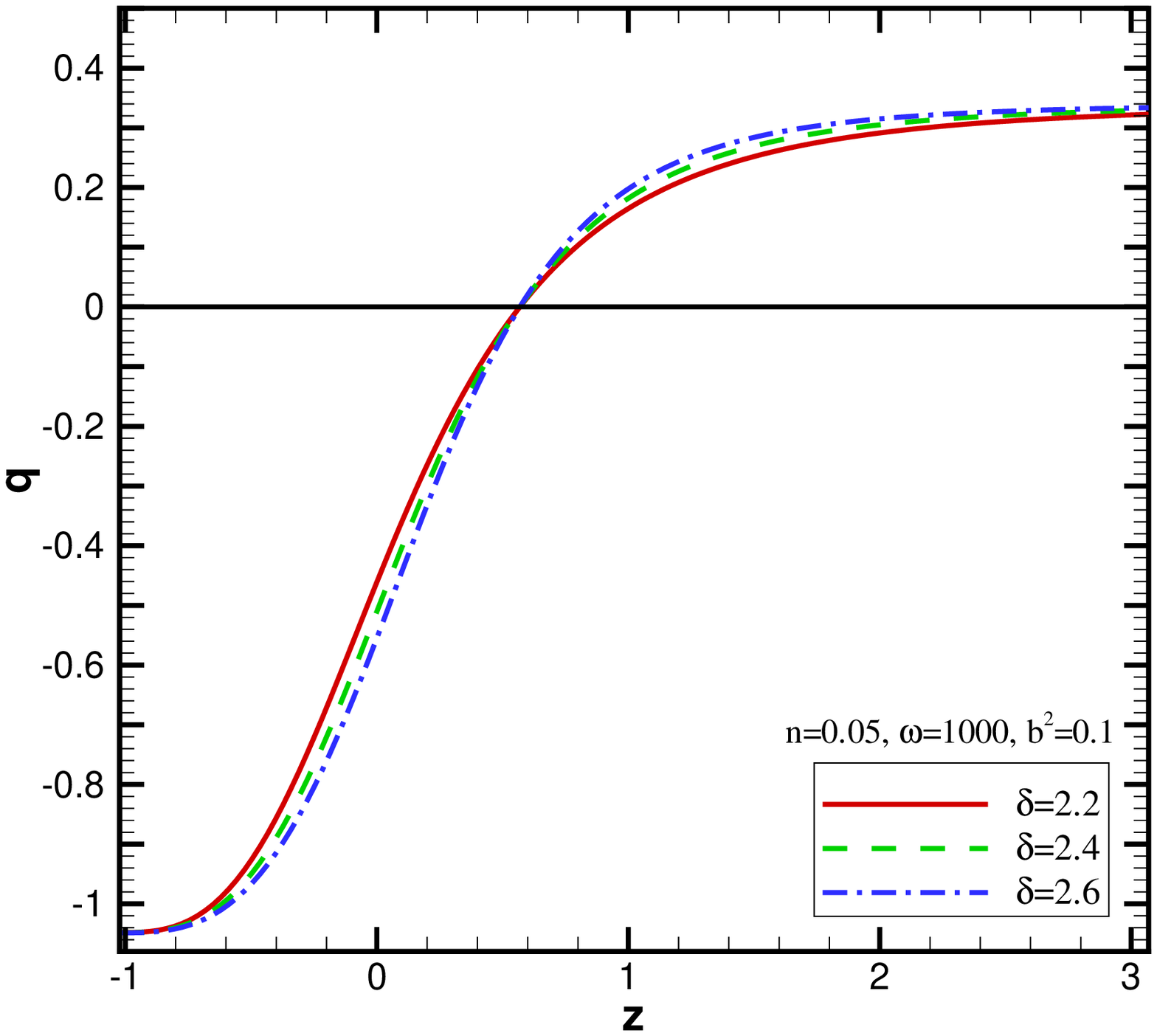}
    \caption{The evolution of $ \omega_D $ and $q$ versus $z$ for the interacting THDE
        in the non-flat BD cosmology.
    Here, we used $ \Omega_{k0}=0.1 $ ,
    $ b^2=0.1 $, $ n=0.05 $ and $\omega=1000$ as the initial conditions.}\label{figq3}
\end{center}
\end{figure}

\begin{figure}[htp]
\begin{center}
 \includegraphics[width=8cm]{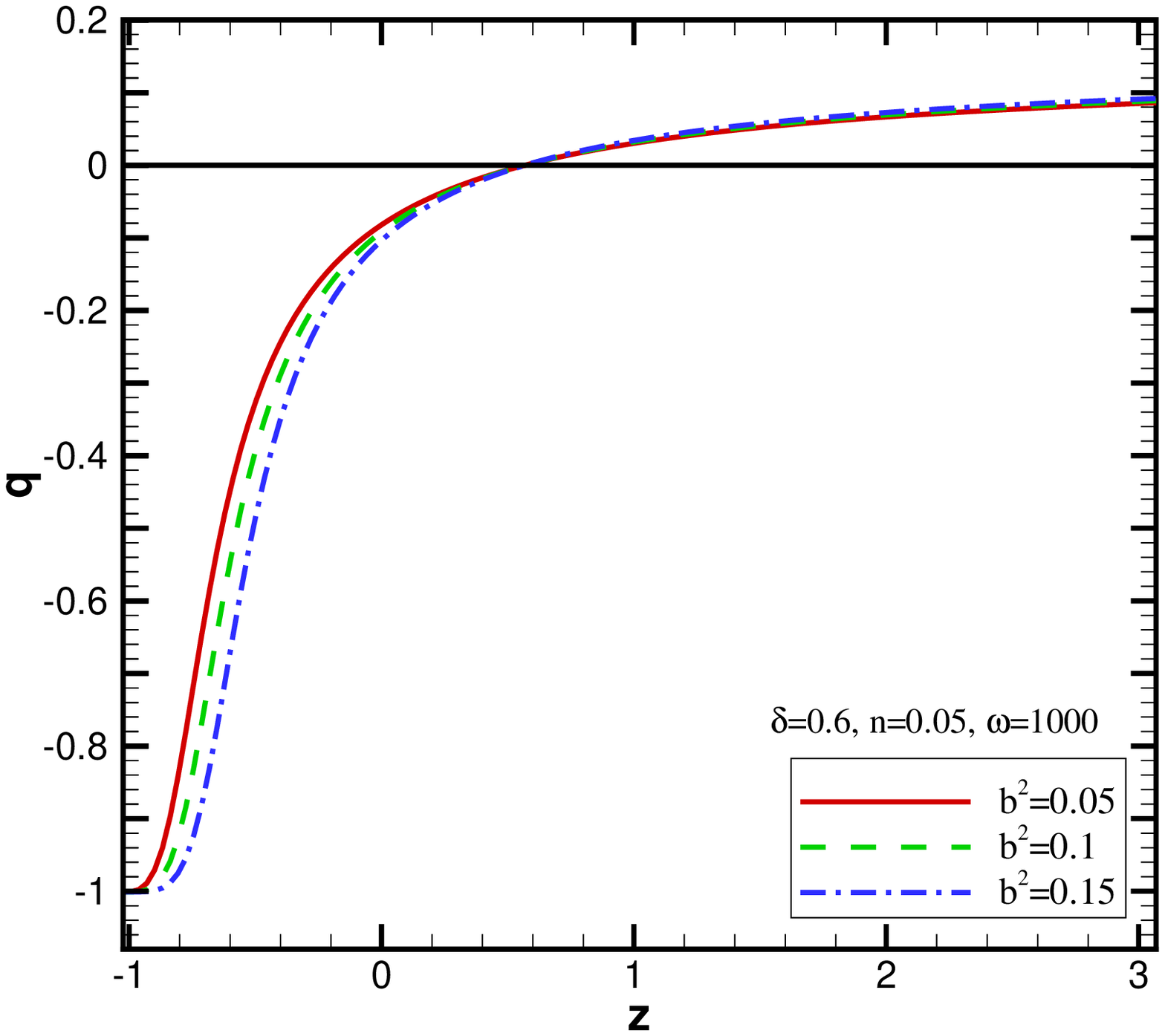}
    \includegraphics[width=8cm]{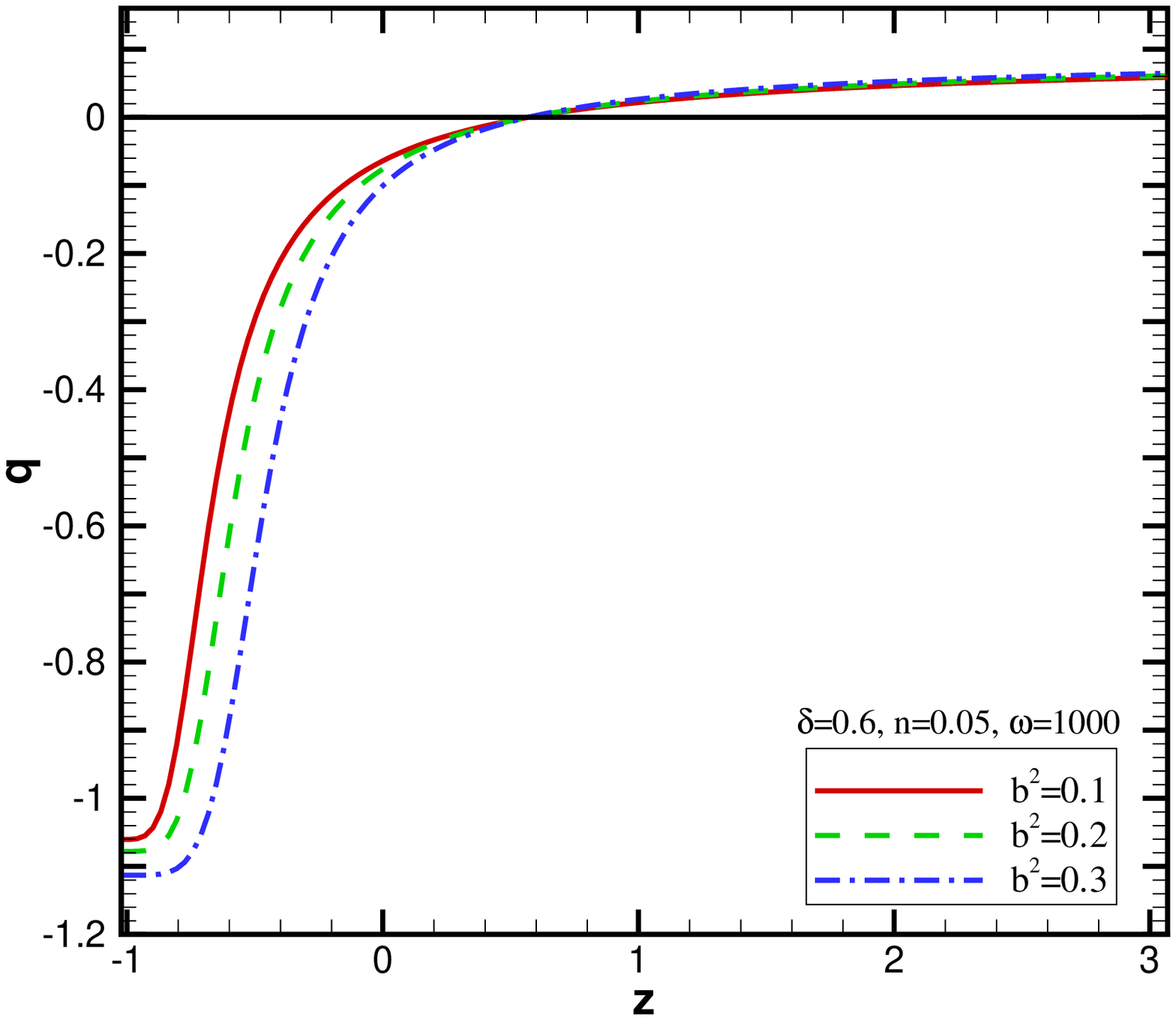}
    \caption{The evolution of $q$ versus $z$ for the interacting THDE.
    We considered $\Omega_k=0 $ (upper panel), $ \Omega_{k0}=0.1 $ (lower panel),
    $ \delta=0.6 $, $ n=0.05 $ and $\omega=1000$ as the initial conditions.}\label{figq2}
    \end{center}
\end{figure}

In the FRW background, filled with DE and DM interacting with each
other, the total energy-momentum conservation law is decomposed into
\begin{equation}
\dot{\rho}_D+3H(1+\omega_D)\rho_D=-Q,\label{QConserveDE}
\end{equation}
and
\begin{equation}
\dot{\rho}_m+3H\rho_m=Q,\label{QConserveCDM}
\end{equation}
where $Q$ denotes the interaction term, and we assume that it has the $Q=3b^2qH(\rho_m+\rho_D)$
form \cite{Wei,Chimento1,Chimento2}, in which $ b^2 $ is the coupling constant.
Taking the time derivative of Eq. (\ref{Friedeq01}) and using Eqs. (\ref{phidot}),
(\ref{phiddot}), (\ref{rhodot}) and (\ref{QConserveCDM}), we have
 \begin{eqnarray}\label{Hdot2}
 &&\frac{\dot{H}}{H^2}=\Big[3\Omega_D-3(1+\Omega_k)(1+b^2)-2\Omega_k(n-1)+\nonumber\\&&\nonumber
 2n(\delta\Omega_D+\frac{2\omega n^2}{3}+(n\omega-3)(b^2+1)-2n-1)\Big]\\&&\nonumber\times
\Big(2(\delta-2)\Omega_D-\frac{4n^2\omega}{3}+(3b^2+2)(2n+1)+\\&& b^2(3\Omega_k-2n^2\omega)\Big)^{-1},
 \end{eqnarray}
substituted into Eq.~(\ref{dotOmega1}) to obtain
 \begin{eqnarray}\label{Omegaprime2}
&&\Omega_D^\prime=2n(\delta-1)\Omega_D+\nonumber\\&&\nonumber\Big[\Omega_D(1-\delta)\Big(3\Omega_D-3(1+\Omega_k)(1+b^2)-2\Omega_k(n-1)
+\\&&\nonumber 2n(\delta\Omega_D+n\omega+\frac{2\omega n^2}{3}+(n\omega-3)(b^2+1)-2n-1))\Big)\Big]\nonumber\\
&&\times\Big((\delta-2)\Omega_D-\frac{2n^2\omega}{3}+(\frac{3b^2}{2}+1)(2n+1)+\\&&\nonumber b^2(3\Omega_k-2n^2\omega)\Big)^{-1}.
 \end{eqnarray}
In the absence of interaction term $ (b^2=0) $, Eq.
(\ref{Omegaprime2}) is reduced to its respective relation in the
previous section. The evolution of $\Omega_D$ against redshift $z$
for interacting THDE has been plotted in Fig. \ref{Omega2}. As it is
seen, we have $ \Omega_D \rightarrow 0 $ and $\Omega_D \rightarrow
1$ at the $z\rightarrow \infty$ (the early time) and $z\rightarrow
-1$ (the the late time) limits, respectively.
\begin{figure}[htp]
\begin{center}
    \includegraphics[width=8cm]{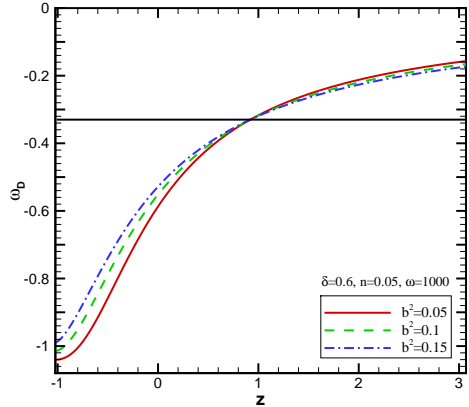}
    \includegraphics[width=8cm]{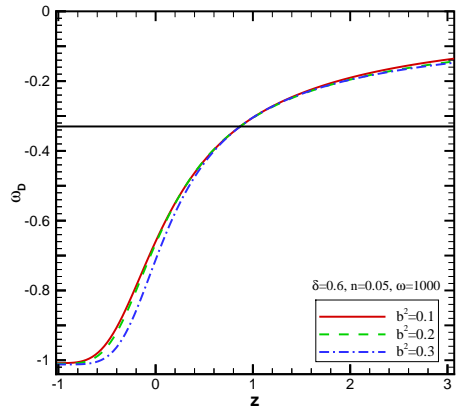}
    \caption{The evolution of $\omega_D$ versus
    $z$ for the interacting THDE, where $\Omega_k=0 $ (upper panel), $\Omega_{k0}=0.1$ (lower panel), $ \delta=0.6 $, $ n=0.05 $ and $\omega=1000$ are adopted as the initial conditions.}\label{figw2}
\end{center}
\end{figure}
The Eos parameter $ \omega_D $ can also be derived by combining Eqs. (\ref{QConserveDE}) and (\ref{rhodot}) with Eq.~(\ref{Hdot2}) as
\begin{eqnarray}\label{w2}
&&\omega_D=-1-\frac{2\delta n}{3}+\frac{(2n-\frac{2\omega n^2}{3}+1)b^2}{\Omega_D}+\nonumber\\
&&\frac{(6n-2\omega n^2+3)b^2+2(\delta-2)\Omega_D}{6\Omega_D}\times\nonumber\\&&\Big[3(\Omega_D-(1+\Omega_k)(1+b^2))-2\Omega_k(n-1)+\nonumber\\
&&2n(\delta\Omega_D+\frac{2\omega n^2}{3}+(n\omega-3)(b^2+1)-2n-1)\Big]\times\nonumber\\
&&\Big(2(\delta-2)\Omega_D-\frac{4n^2\omega}{3}+(3b^2+2)(2n+1)+\nonumber\\&& b^2(3\Omega_k-2n^2\omega)2\omega\Big)^{-1}.
\end{eqnarray}
We have also plotted the evolution of $ \omega_D $ versus $ z $ for
the interacting THDE in Figs. \ref{figw2} and \ref{figq3} for for
$n=0.05$ and $\omega=1000$ \cite{jcap}. From these figures, it is
obvious that, depending on the values of $\delta$, $\Omega_k$ and
$b^2$, the phantom line can be crossed, and the cosmological
constant model of DE ($\omega_D\rightarrow-1$) is obtainable at the
$z\rightarrow-1$ limit in both the flat (for $0.5<\delta<1$ and
$b^2>0.1$) and non-flat (for $\delta>1$) FRW universes. From Eq.
(\ref{Hdot2}), we also get
\begin{eqnarray}\label{q2}
&&q=-1-\Big[3(\Omega_D-(1+\Omega_k)(1+b^2))-2\Omega_k(n-1)+\nonumber\\&&\nonumber
2n(\delta\Omega_D+\frac{2\omega n^2}{3}+(n\omega-3)(b^2+1)-2n-1)\Big]\times\\&&\nonumber
\Big(2(\delta-2)\Omega_D-\frac{4n^2\omega}{3}+(3b^2+2)(2n+1)+\nonumber\\&& b^2(3\Omega_k-2n^2\omega)2\omega\Big)^{-1}.
\end{eqnarray}
It is obvious that, in the limiting case $ b=0 $, the respective
relation in the previous section can be retrieved. The evolution of
$ q $ versus $ z $ has been plotted in Figs. \ref{figq3} and
\ref{figq2}.

From Figs. \ref{figq3} and \ref{figq2}, it is clear that $ q $
starts from positive value at the earlier time, and takes the
negative values later, and also, it has a zero at $ z\approx 0.6 $
\cite{Daly,Komatsu,Salvatelli}. Figs. \ref{Omega2}-\ref{figq2}
indicate that, with the same set of the system parameters
($\delta,n,\omega,b$), acceptable and proper behavior for
$\omega_D$, $q$ and $\Omega_D$ is obtainable simultaneously only in
the non-flat universe. In fact, as the non-interacting case, the
non-flat universe can produce more better and acceptable results
compared with the flat universe.
\section{Stability}

In this section we would like to study the classical stability of the obtained models against perturbations.
In the perturbation theory, an important quantity is the squared of the sound speed $ v_s^2 $.
Stability or instability of a given perturbation in the background, can be specified by determining the sign of $ v_s^2 $.
For $ v_s^2>0 $ the given perturbation propagates in the environment meaning that the model is stable against the perturbations.

The squared sound speed $ v_s^2 $ is given by
\begin{equation}\label{v_s}
v_s^2=\frac{dp}{d\rho_D}=\frac{\dot{p}}{\dot{\rho}_D}.
\end{equation}
By differentiating of $ p_{D} $ with respect to time, inserting the
result in Eq. (\ref{v_s}), and using Eq. (\ref{rhodot}), we can
finally get
\begin{equation}\label{v1}
v_s^2=\omega_D+\frac{\omega_D^\prime}{2\delta n+2(2-\delta)\frac{\dot{H}}{H^2}}.
\end{equation}
\begin{figure}[htp]
\begin{center}
      \includegraphics[width=8cm]{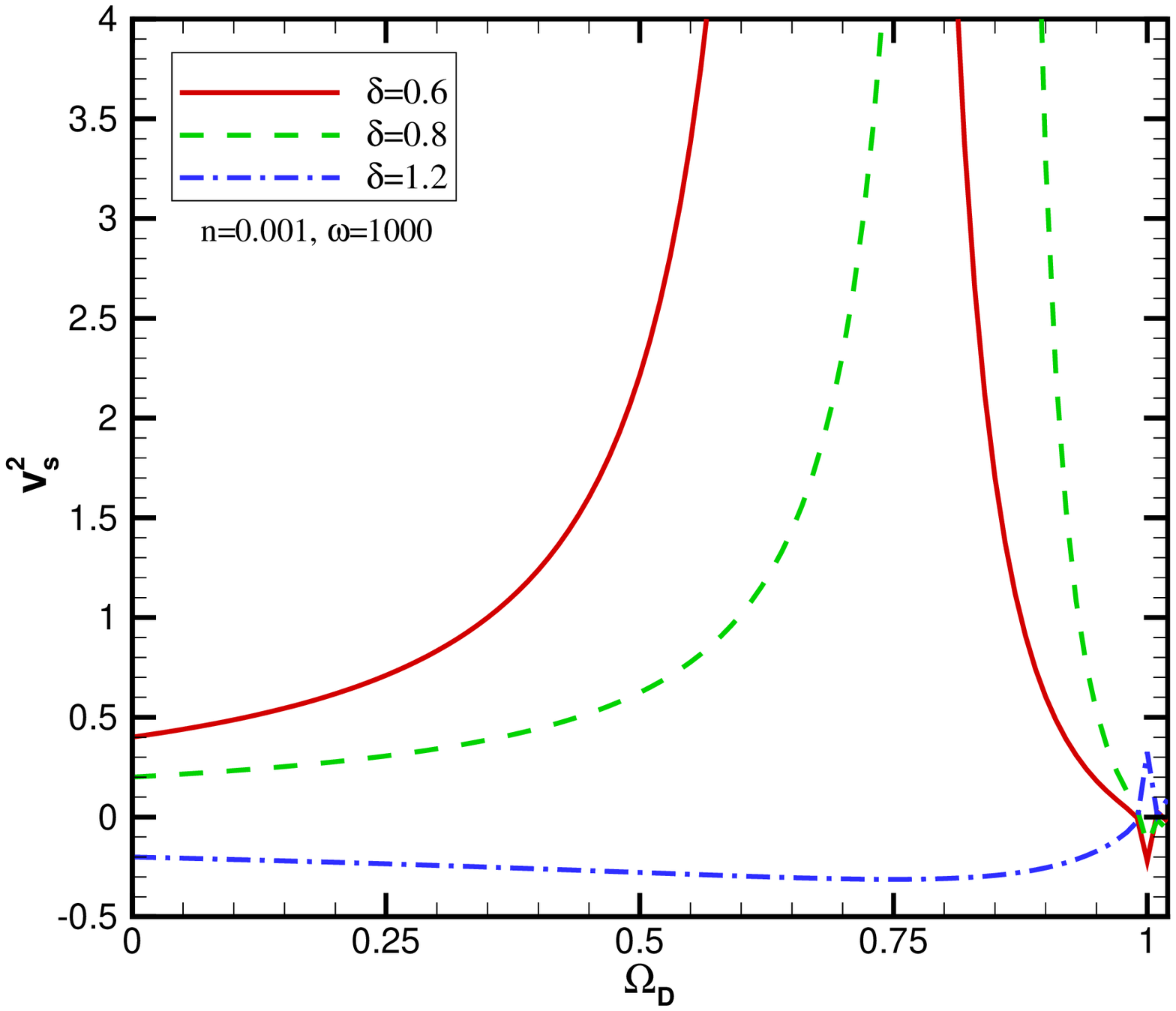}
      \includegraphics[width=8cm]{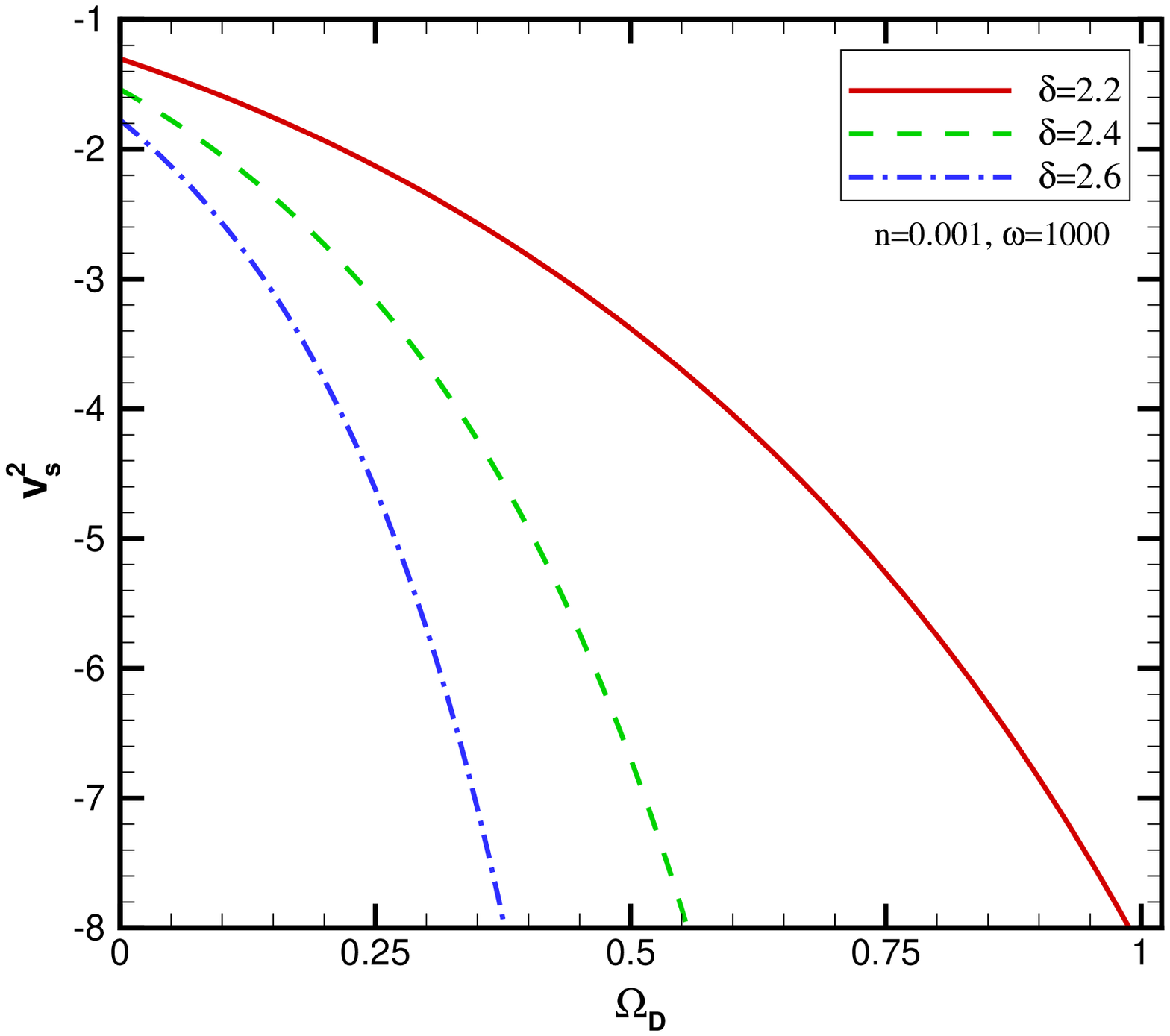}
      \caption{$v_s^2$ versus $z$ for the non-interacting THDE in the BD gravity, where $\Omega_k=0 $ (upper panel), $ \Omega_{k0}=0.1 $ (lower panel), $ n=0.001 $ and $\omega=1000$ \cite{jcap} are adopted.}\label{figV1}
\end{center}
\end{figure}
\subsection{Non-interacting case}
Taking the time derivative of Eq. (\ref{w1}) and using Eqs.
(\ref{rhodot}), (\ref{Hdot1}), (\ref{Omegaprime1}) and (\ref{v1}),
one can obtain $ v_s^2 $ for the non-interacting THDE with the
Hubble cutoff in the BD cosmology. Since this expression is too
long, we shall not present it here, and only plot it in Fig.
\ref{figV1}. This figure shows that, in the flat FRW universe, the
non-interacting THDE model is classically stable (unstable) for $
0<\delta<1 $ ($\delta>1$). In addition, the lower panel indicates
that the model is classically unstable in the non-flat FRW universe.
\subsection{Interacting case}
\begin{figure}[htp]
\begin{center}
     \includegraphics[width=8cm]{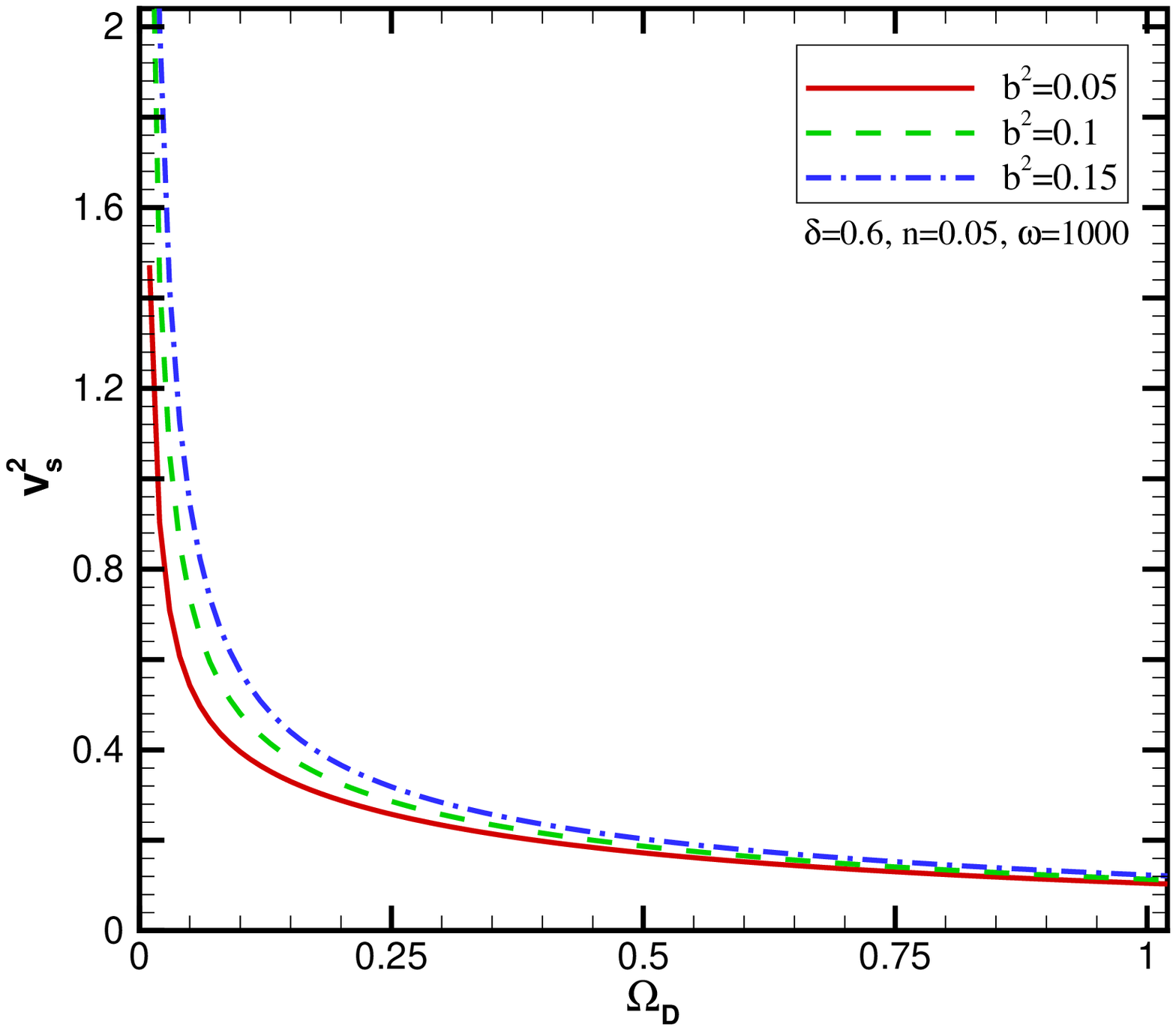}
     \includegraphics[width=8cm]{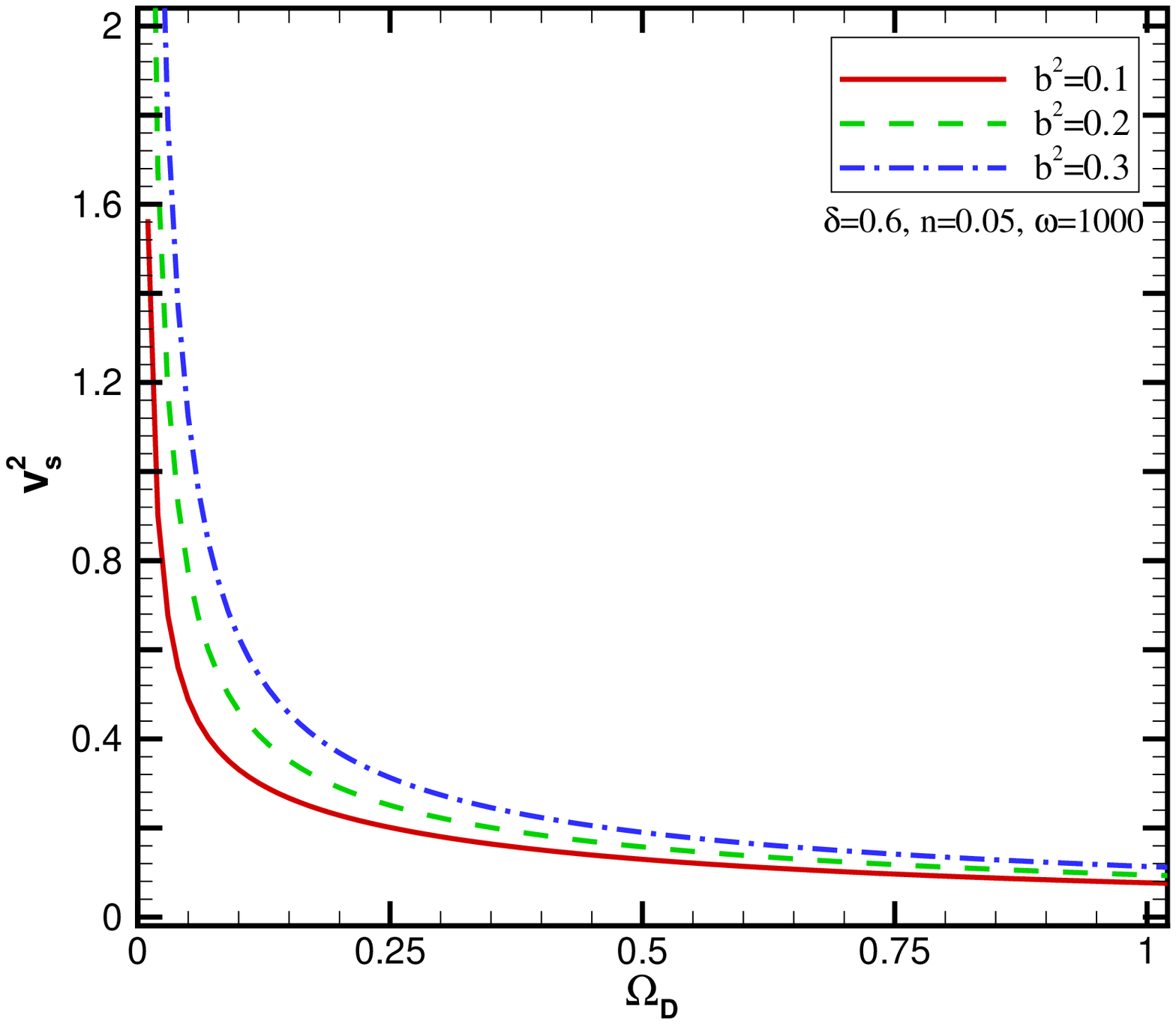}
     \caption{$v_s^2$ versus
     $\Omega_D$ for the sign-changeable interacting THDE
     with the Hubble cutoff. Here, we used $\Omega_k=0 $ (upper panel), $ \Omega_{k0}=0.1 $ (lower panel), $ \delta=0.6 $, $ n=0.05 $ and $\omega=1000$ as the initial conditions.}\label{figV2}
\end{center}
\end{figure}
\begin{figure}[htp]
\begin{center}
    \includegraphics[width=8cm]{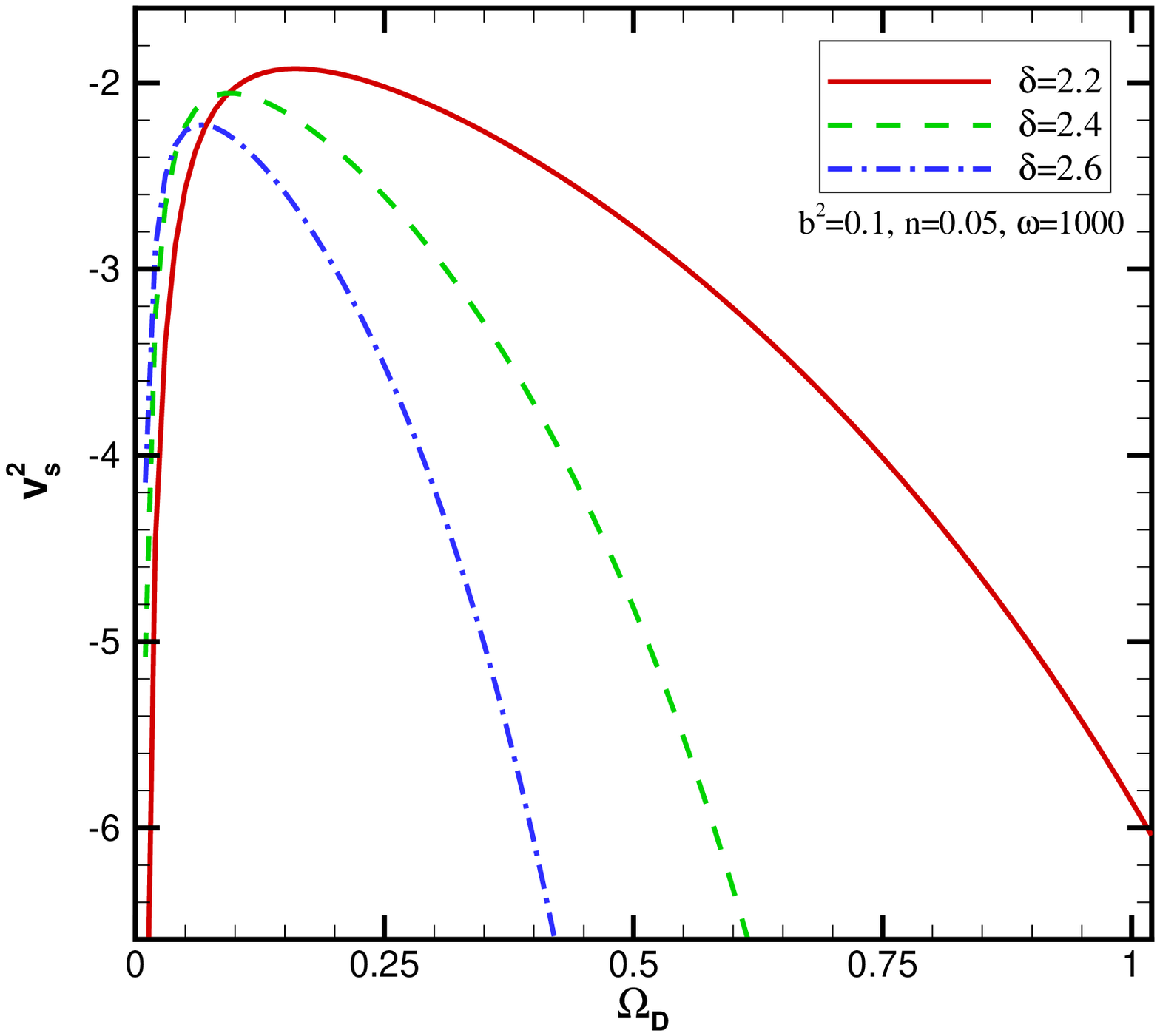}
    \caption{$v_s^2$ versus
    $\Omega_D$ for the sign-changeable interacting THDE
    with the Hubble cutoff, where $ \Omega_{k0}=0.1 $ , $ b^2=0.1 $, $ n=0.05 $ and $\omega=1000$ are adopted.}\label{figV3}
\end{center}
\end{figure}
By taking the time derivative of Eq. (\ref{w2}), and combining the
result with Eqs. (\ref{rho1}), (\ref{rhodot}), (\ref{Hdot2}) and
(\ref{v1}), we can obtain $v_s^2$ for the interacting THDE. Again,
since this expression is too long, we do not present it here, and it
has been plotted in Figs. \ref{figV2} and \ref{figV3}. As the upper
panel of Fig. \ref{figV2} shows, the model is classically stable in
the flat universe, but the values chosen for $\delta$, $b^2$, $n$
and $\omega$ cannot produce proper behavior for $q$, $\omega_D$ and
$\Omega_D$ simultaneously. In fact, in a flat FRW universe, the
interacting THDE cannot produce stable and acceptable description
for $q$, $\omega_D$ and $\Omega_D$ with the same set of
($\delta,n,\omega,b$) simultaneously. The same story is also
obtained in the non-flat case. As it is obvious from Fig.
\ref{figV3} and the lower panel of Fig. \ref{figV2}, the model
description of the cosmic evolution may be stable, depending on the
value of $\delta$. Indeed, although, the parameters leading to the
stable description provide suitable behavior for $q$ and $\omega_D$,
they cannot produce proper behavior for $\Omega_D$.

\section{Concluding remarks}
We studied the consequences of using THDE in order to model
\textbf{DE} in the BD framework. For the flat universe and the
non-interacting THDE, it has been obtained that, for the assumed
initial conditions, only the $\delta=1.2$ case can produce suitable
behavior for $q$, $\omega_D$ and $\Omega_D$. For this case, we
obtained $\omega_D(z\rightarrow-1)\rightarrow-1$ addressing us that
this model simulates cosmological constant at future. The classical
stability analysis also shows that this model is not stable. If the
interaction $Q=3b^2qH(\rho_m+\rho_D)$ \cite{Wei,Chimento1,Chimento2}
is added to the system, then there is not a set of
($\delta,n,\omega,b$) leading to acceptable and proper behavior for
$q$, $\omega_D$ and $\Omega_D$ simultaneously meaning that the
system description is incomplete. It is also useful to mention here
that this incomplete description can meet the classical stability
requirement.

For the non-flat FRW universe, we found out that $i$) for
$\delta>2$, the model can provide suitable descriptions for the
cosmic evolution in both the interacting and non-interacting cases,
$ii$) these descriptions are not stable, and $iii$) there are cases
for which $\omega_D(z\rightarrow-1)\rightarrow-1$ the same as the
EoS of the cosmological constant ($\omega_D=-1$). In fact, the same
as the flat universe, the sets of ($\delta,n,\omega,b$) leading to
the stable cases cannot provide proper explanations for $q$,
$\omega_D$ and $\Omega_D$ simultaneously and vice versa.

\acknowledgments{The work of H. Moradpour has been supported
financially by Research Institute for Astronomy and Astrophysics of
Maragha (RIAAM) under project No. 1/5750-8. JPMG and IPL thank
Coordena\c c\~ao de Aperfei\c coamento de Pessoal de N\'ivel
Superior (CAPES-Brazil), VBB thanks Conselho Nacional de
Desenvolvimento Cient\'ifico e Tecnol\'ogico (CNPq-Brazil).}

\end{document}